\documentclass[aps,twocolumn,showpacs,preprintnumbers,prb,amsmath,amssymb,amsfonts,superscriptaddress,floatfix]{revtex4-1}

\usepackage[latin1]{inputenc}
\usepackage{graphics}
\usepackage{color}
\usepackage{amssymb}
\usepackage{amsmath}
\usepackage{overpic}
\usepackage{epstopdf}
\usepackage{bm}
\usepackage{graphicx}
\usepackage{subfigure}

\begin{document}

\title{Structural Instability and Magnetism of Superconducting KCr$_3$As$_3$}

\author{Guangzong Xing}
\thanks{These authors contributed equally}
\affiliation{Department of Physics and Astronomy, University of Missouri,
Columbia, MO 65211-7010, USA}
\affiliation{College of Materials Science and Engineering,
Jilin University, 130012, Changchun, China}

\author{Ling Shang}
\thanks{These authors contributed equally}
\affiliation{Department of Physics, Shanghai University, Shanghai 200444,
China}

\author{Yuhao Fu}
\affiliation{Department of Physics and Astronomy, University of Missouri,
Columbia, MO 65211-7010, USA}

\author{Wei Ren}
\email{renwei@shu.edu.cn}
\affiliation{Department of Physics,
Shanghai Key Laboratory of High Temperature Superconductors,
MGI and ICQMS, Shanghai University, Shanghai 200444,
China}

\author{Xiaofeng Fan}
\email{xffan@jlu.edu.cn}
\affiliation{College of Materials Science and Engineering,
Jilin University, 130012, Changchun, China}

\author{Weitao Zheng}
\email{wtzheng@jlu.edu.cn}
\affiliation{College of Materials Science and Engineering,
Jilin University, 130012, Changchun, China}

\author{David J. Singh}
\email{singhdj@missouri.edu}
\affiliation{Department of Physics and Astronomy, University of Missouri,
Columbia, MO 65211-7010, USA}

\date{\today}

\begin{abstract}
We find a lattice instability in the superconductor KCr$_3$As$_3$,
corresponding to a distortion of the Cr metallic wires in the crystal structure.
This distortion couples strongly to both the electronic and magnetic
properties, in particular by making the electronic structure much more
nearly one-dimensional, and by shifting the compound away from magnetism.
We discuss the implications of these results in the context of the possibly
unconventional superconductivity of this phase.
\end{abstract}

\maketitle

\section{Introduction}

The superconductivity of the Cr-based materials,
exemplified by K$_2$Cr$_3$As$_3$ and KCr$_3$As$_3$,
has attracted recent interest. 
\cite{Na2Cr3As3,KCr3As3-5K,K2Cr3As3,Rb2Cr3As3,Rb2Cr3As3-1,Rb2Cr3As3-2}
This arises in part from the possible connections with the Fe-based
superconductors, and from unique features of these materials. These
include, non-centrosymmetry,
proximity to magnetism and quasi-one-dimensional
structural features in the crystal
structure. A variety of related
compounds have been found, including both Cr and Mo compounds.
The structures of these materials,
$A_xM_3$As$_3$ (with $A$ = Na, K, Rb, Cs, $M$=Cr, Mo)
contain metal-arsenic tubes
separated by alkali-metal cations.
\cite{Na2Cr3As3,K2Cr3As3-Rb,K2Cr3As3-Rb-Cs,K2Cr3As3-1,K2Cr3As3-2,
K2Cr3As3-3,Cs2Cr3As3-1,Cs2Cr3As3,K2Mo3As3,Cs2Mo3As3-Rb}
The superconducting transition temperature
of K$_2$Cr$_3$As$_3$ is $T_c$=6.1 K,\cite{K2Cr3As3} and decreases with
pressure.
This has been associated with
changes in metal-pnictogen bond angles analogous to the Fe-based
pnictide superconductors.\cite{K2Cr3As3-press}
These compounds occur in hexagonal crystal structures, as mentioned, with
Cr (or Mo) wires running along the $c$-axis direction. These wires consist
of perfect Cr$_3$ triangles stacked along $c$, and coordinated on the
outside by As.

Theoretical studies of K$_2$Cr$_3$As$_3$ show a complex Fermi surface structure
with three-dimensional and one-dimensional sheets.
\cite{wu,subedi}
The one-dimensional sheets have been seen in angle resolved photoemission
(ARPES) experiments.
\cite{watson}
There is also a suppression of spectral weight near the
Fermi energy following Tomonaga-Luttinger liquid behavior, characteristic
of a one-dimensional electronic system.
The electronic structure near the Fermi energy
is mainly from the Cr-3d electrons.\cite{K2Cr3As3-cal,wu,subedi}
Theoretical calculations find weak magnetic instabilities at the
zone center and competing magnetic states. \cite{wu}
This would be consistent
with triplet superconducting state, following similar arguments to
Sr$_2$RuO$_4$. \cite{rice,machida,mazin}
This magnetic tendency is a consequence of the high electronic
density of states at the Fermi level, $N(E_F)$, associated with the
multiple Fermi surface sheets.\cite{K2Cr3As3-cal,wu}
A triplet state was additionally suggested based on a Hubbard model
for the Cr wires. \cite{zhong}
Nearness to ferromagnetism has been supported by NMR measurements
in Rb$_2$Cr$_3$As$_3$ and K$_2$Cr$_3$As$_3$, which
find an increasing spin-lattice relaxation rate with cooling.
The Hebel-Slichter peak is absent, suggesting the possibility
of an unconventional superconducting state.
\cite{K2Cr3As3-3,Rb2Cr3As3-2}
However, lack of a Hebel-Slichter peak may also be a consequence
of strong coupling in a conventional scenario, such as that proposed by
Subedi. \cite{subedi}
In any case, spin fluctuations are also seen in neutron scattering data.
However, they are much weaker than in, e.g. Fe-based superconductors,
and are not at the zone center.
\cite{taddei}
Muon spin relaxation ($\mu$SR) experiments are reported to be consistent with
an unconventional singlet state.
\cite{K2Cr3As3-1,Cs2Cr3As3-1}
However, it was noted that this
may also be consistent
with a fully gapped conventional $s$ wave state.
\cite{K2Cr3As3-1}
In contrast, penetration depth measurements suggest nodes in the gap.
\cite{pang}
However,
the situation is complicated by the non-centrosymmetric crystal structure of
K$_2$Cr$_3$As$_3$, which can mix singlet and triplet states.

Here we focus on the KCr$_3$As$_3$ compound, which has a 
somewhat lower electron
count in the Cr$_3$As$_3$ tubes, but is otherwise reported to be very
similar in superconducting properties to compounds in the K$_2$Cr$_3$As$_3$
family.
\cite{KCr3As3,KCr3As3-1,KCr3As3-5K}
Similar to K$_2$Cr$_3$As$_3$,
the ground state reported by theoretical calculations is magnetic, and
the Fermi surface shows one-dimensional and three-dimensional sheets.
\cite{KCr3As3-cal}
However, magnetic order is not seen in experiments.
Interestingly though, Bao and co-workers reported a possible spin-glass
ground state in non-superconducting samples. This was accompanied
by a Curie-Weiss
behavior of the susceptibility with a fitted effective moment of 0.68
$\mu_B$. \cite{KCr3As3}
In contrast, experiments on other samples find different behavior,
\cite{KCr3As3-5K} perhaps related to subtle chemical differences due to 
different sample preparations.
Replacement of K by Rb leads to an increase in critical temperature
from $\sim$5 K to 7.3 K, showing that the superconductivity in this
family is as robust as that in the K$_2$Cr$_3$As$_3$ compounds.
\cite{liu}
The compound is reported to occur in a hexagonal TlFe$_3$Te$_3$-type
crystal structure, spacegroup $P6_3/m$ (number 176),
which is centrosymmetric.
\cite{KCr3As3,KCr3As3-5K}

Here we show that hexagonal KCr$_3$As$_3$ has a lattice instability.
This is related to recent experimental and theoretical findings for
hexagonal K$_2$Cr$_3$As$_3$. \cite{K2Cr3As3-new}
This has strong effects on the magnetic and electronic
properties, with implications for the superconductivity.

\section{Methods}

The reported results were obtained within density functional theory (DFT)
using the Perdew-Burke-Ernzerhof generalized gradient approximation
(PBE GGA), \cite{vaspgga}
and the experimentally reported lattice parameters,
$a$=9.0909 \AA, $c$=4.1806 \AA. The internal atomic
coordinates are relaxed by total energy minimization, subject to symmetry.
The shortest Cr-Cr bond length in a triangular layer is 2.58 \AA, which
is almost the same as the distance of 2.57 \AA, between closest Cr in adjacent
layers with the hexagonal structure.

Phonon calculations were done using the finite difference
supercell method using the PHONOPY code. \cite{phonopy}
We used 2$\times$2$\times$4 supercells with underlying
DFT calculations performed using the projector-augmented wave
method as implemented in the VASP code. \cite{vasppaw,vaspcode}
The energy cutoff was 500 eV and 
a Brillouin zone integration grid
spacing of 2$\pi\times$0.032 \AA$^{-1}$ was used.
A phonon instability was found at the $\Gamma$ point.
We relaxed the crystal structure allowing the condensation
of the unstable modes, first of all using VASP, and then performed a
final relaxation using the general potential
linearized augmented planewave (LAPW) method,
\cite{wien2klapw}
as implemented in WIEN2k. \cite{wien2kcode}
The LAPW method was used for all the electronic structure and magnetic
calculations and for the energies reported here.
These calculations were done using well converged basis sets
obtained with an interstitial planewave sector cutoff set by
$RK_{max}$=9, where $R$ is the smallest LAPW sphere radius and $K_{max}$
is the cutoff. Local orbitals were used for the semicore states.
Structure relaxations and energies
were obtained including relativity at the scalar relativistic level.
Spin orbit was included in the reported electronic structures, and dense
converged samplings of the Brillouin zone were employed.

\section{Results and Discussion}

\begin{figure}
\includegraphics[width=0.70\columnwidth,angle=0]{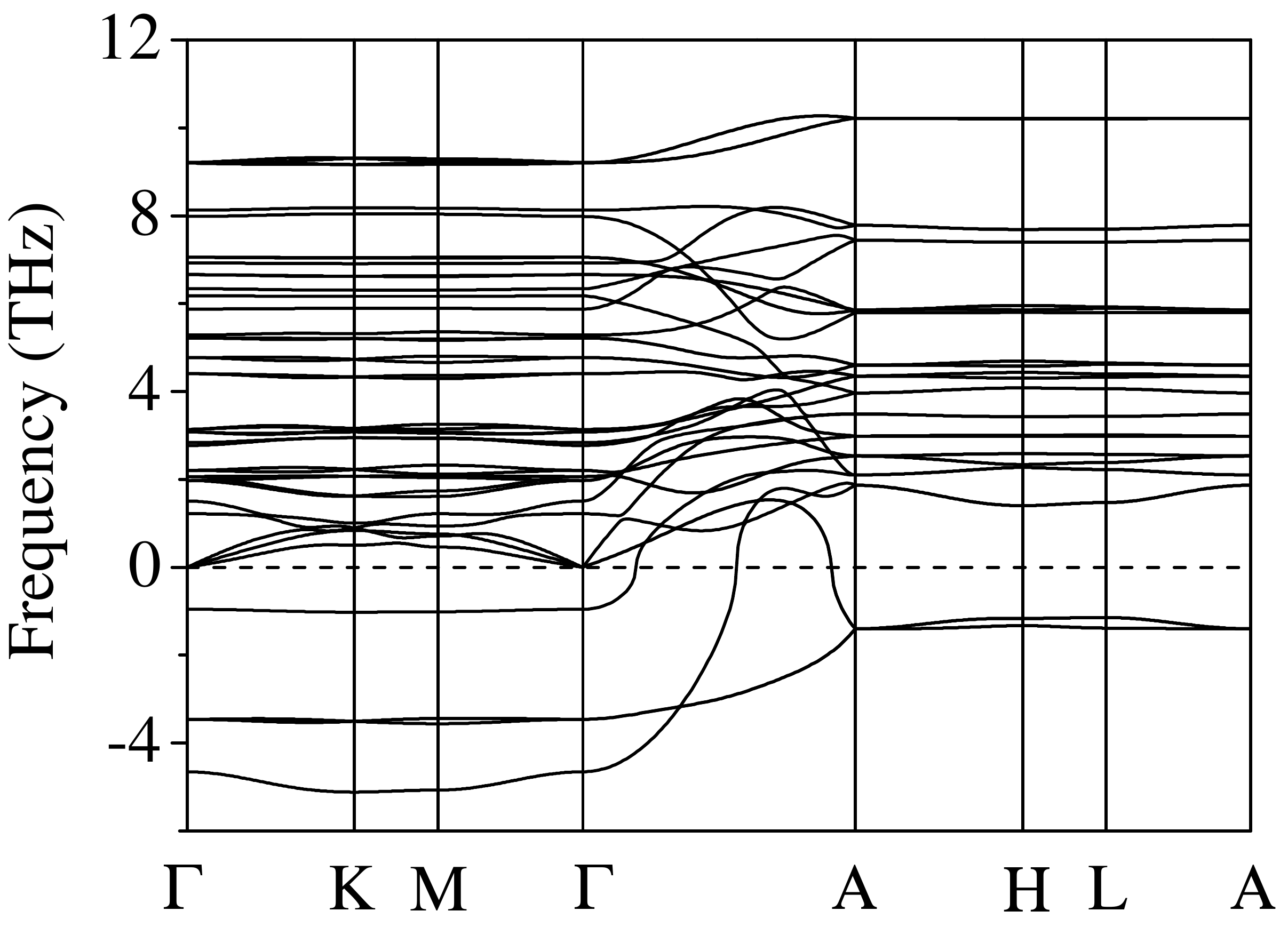}
\caption{Calculated phonon dispersions of KCr$_3$As$_3$ in
its experimentally observed hexagonal structure.
Note the unstable branches with imaginary
frequencies shown below the horizontal axis.}
\label{phonons}
\end{figure}

Our calculated phonon dispersions for the
experimentally observed hexagonal structure are shown in Fig. \ref{phonons}.
We find very strong phonon instabilities in the $k_z$=0
plane with four unstable branches, and weaker instabilities at $k_z$=1/2.
These branches show little dispersion at constant $k_z$.
This is qualitatively similar to K$_2$Cr$_3$As$_3$, \cite{K2Cr3As3-new}
but the instabilities
in the present case are stronger and there are more unstable branches.
This and the weak dispersion of the modes in the basal ($k_z$=0) plane
imply a strong structural instability of the individual
Cr$_3$As$_3$ tubes are consistent
with little correlation between tubes.

\begin{figure}
\includegraphics[width=\columnwidth,angle=0]{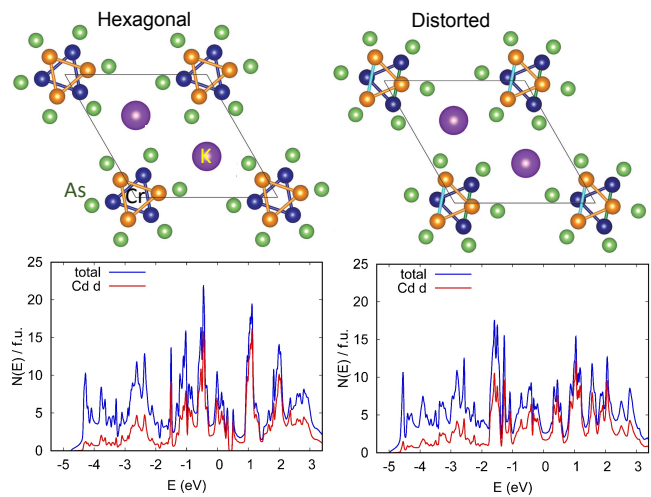}
\caption{(top) Structure of undistorted KCr$_3$As$_3$, with the Cr and
Cr-Cr bonds in
the two layers shown in different colors; (b) distorted KCr$_3$As$_3$
with different bond lengths shown with different colors.
(bottom) Electronic density of states on a per formula unit basis
for the two structures, showing the Cr $d$ projections. The Fermi
energy is at 0 eV.}
\label{struct}
\end{figure}

\begin{figure}
\includegraphics[width=0.70\columnwidth,angle=0]{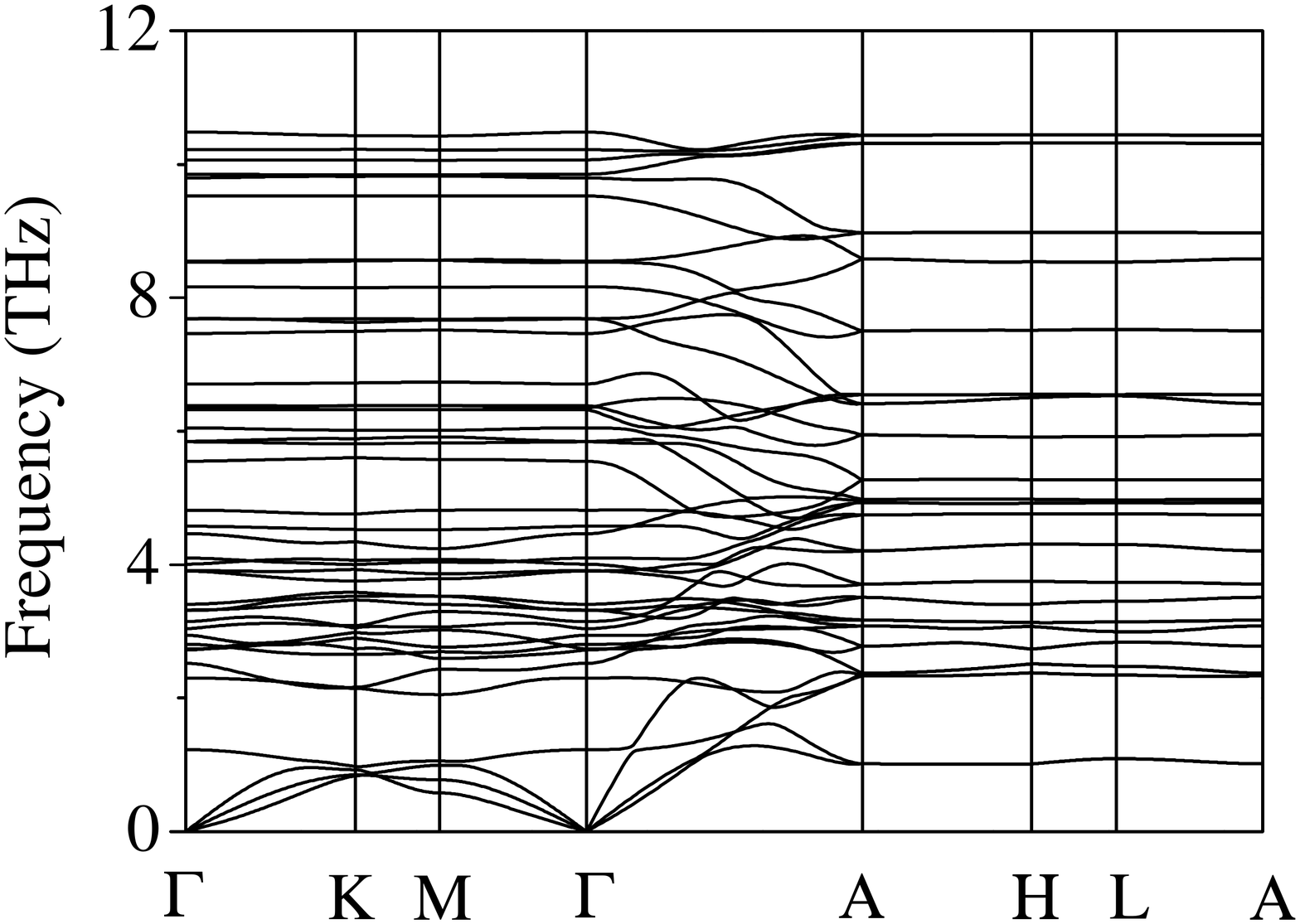}
\caption{Calculated phonon dispersions of KCr$_3$As$_3$ with the
distorted structure. The labels follow those of the ideal hexagonal structure
for ease of comparison,
although these are not proper symmetry points for the monoclinic zone.
Note that no unstable branches are present.}
\label{phonon-dist}
\end{figure}

We explored this structural instability by fully relaxing the atomic
coordinates without symmetry, starting from a displacement corresponding
to the most unstable phonon at $\Gamma$. This relaxation was done
within a single unit cell to explore the effect of the distortion of the
Cr$_3$As$_3$ tubes. The actual structure likely has disorder between
different tubes similar to K$_2$Cr$_3$As$_3$.
\cite{K2Cr3As3-new}
This is consistent with low dispersion of the unstable modes in the $k_z$=0
plane. In particular, flat dispersions mean that low energy differences between
different in plane orderings (i.e. between different tubes) are to be expected
with a resulting low coherence length for this type of distortion, even though
the distortion energy is large with respect to the ideal hexagonal structure.

The relaxation yielded to a structure with spacegroup $P$2$_1$/$m$.
The distortion consists primarily of a distortion of the Cr$_3$ wires
in the crystal structure and is depicted in Fig. \ref{struct}.
Compared with the reported structure with spacegroup $P$6$_3$/$m$,
it shows three different types of Cr atom and two different Cr-Cr bonds
bond lengths, 2.74 \AA\ and 2.42 \AA\, respectively.
The energy gain from the distortion is 0.15 eV per formula unit.

This is a very sizable energy that is well outside the range expected for
DFT errors.
It also is much larger than the magnetic energies for the ideal structure
(see below).
This means that the ideal structure cannot be stabilized by
magnetism.
We also performed phonon calculations for the distorted structure.
The resulting phonon dispersion is shown in Fig. \ref{phonon-dist}.
As seen, the distorted structure does not exhibit unstable modes.
Significantly, centrosymmetry is retained in the distorted structure.

\begin{figure}
\includegraphics[width=0.80\columnwidth,angle=0]{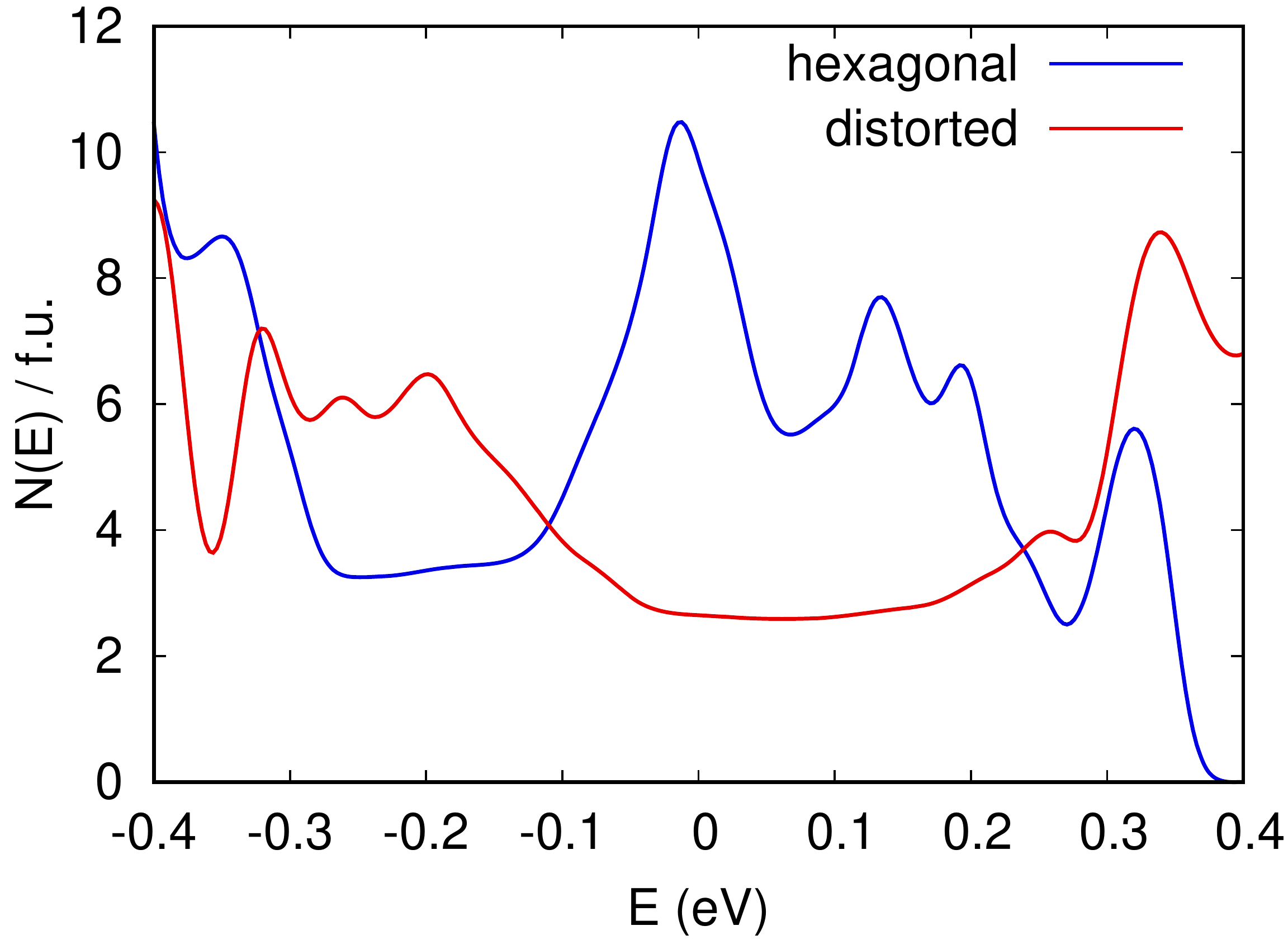}
\caption{Comparison of the electronic density of states for hexagonal
and distorted KCr$_3$As$_3$ near the Fermi energy. Note that the
distortion strongly depletes $N(E_F)$.}
\label{dos-comp}
\end{figure}

The electronic density of states (DOS) are
shown for the two structures in lower panels of Fig. \ref{struct}.
The DOS is mainly derived from Cr $d$
states in the region near the Fermi energy, $E_F$.
The details of the DOS are
changed by the distortion, but the overall shape
of the DOS is similar for the distorted and
undistorted structures.
However, as shown in Fig. \ref{dos-comp},
there is a clear depletion in the DOS near $E_F$ in the
distorted structure associated in the removal of a peak at $E_F$ for the
undistorted structure.
At $E_F$, the DOS, $N(E_F)$=9.9 eV$^{-1}$ per formula unit for
the ideal hexagonal structure, but falls to 
$N(E_F)$=2.6 eV$^{-1}$ for the distorted monoclinic structure.
The Cr $d$ contribution as determined by projection onto the LAPW
sphere, radius 2.25 bohr, is 6.5 eV$^{-1}$ for the hexagonal structure,
which amounts to $\sim$2.2 eV$^{-1}$ per atom (both spins). This is 
slightly above
the expected critical value for meeting the Stoner criterion
for itinerant magnetism. \cite{janak}
The distorted structure
with its much lower Cr $d$ contribution of $\sim$0.6 eV$^{-1}$ per Cr
is far from this condition. Thus the distortion may be expected to strongly
affect the magnetic properties.

\begin{figure}
\includegraphics[width=0.85\columnwidth,angle=0]{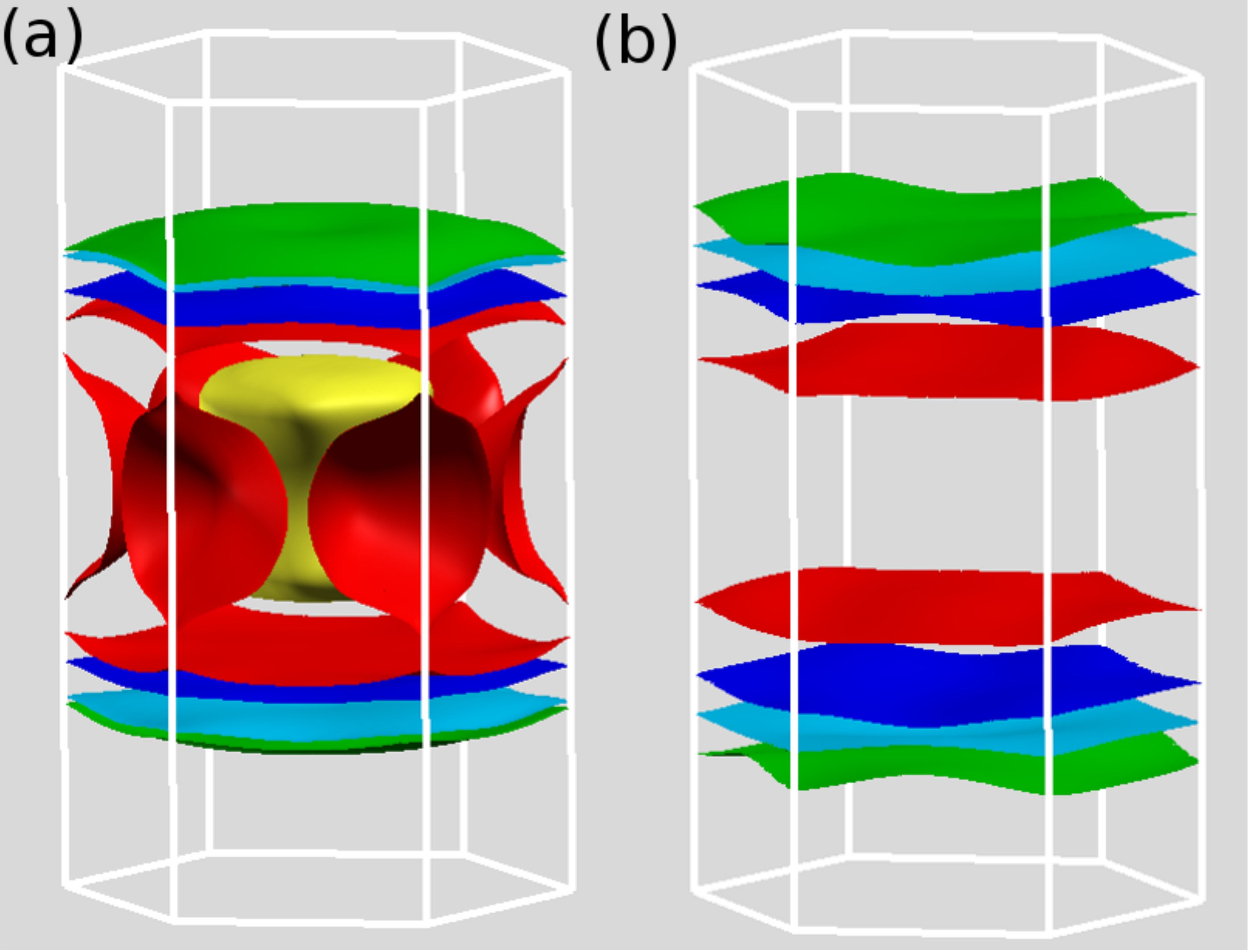}
\caption{Fermi surfaces for the undistorted (a, left) and distorted (b, right)
structures of KCr$_3$As$_3$.}
\label{fermi}
\end{figure}

\begin{figure}
\includegraphics[width=0.95\columnwidth,angle=0]{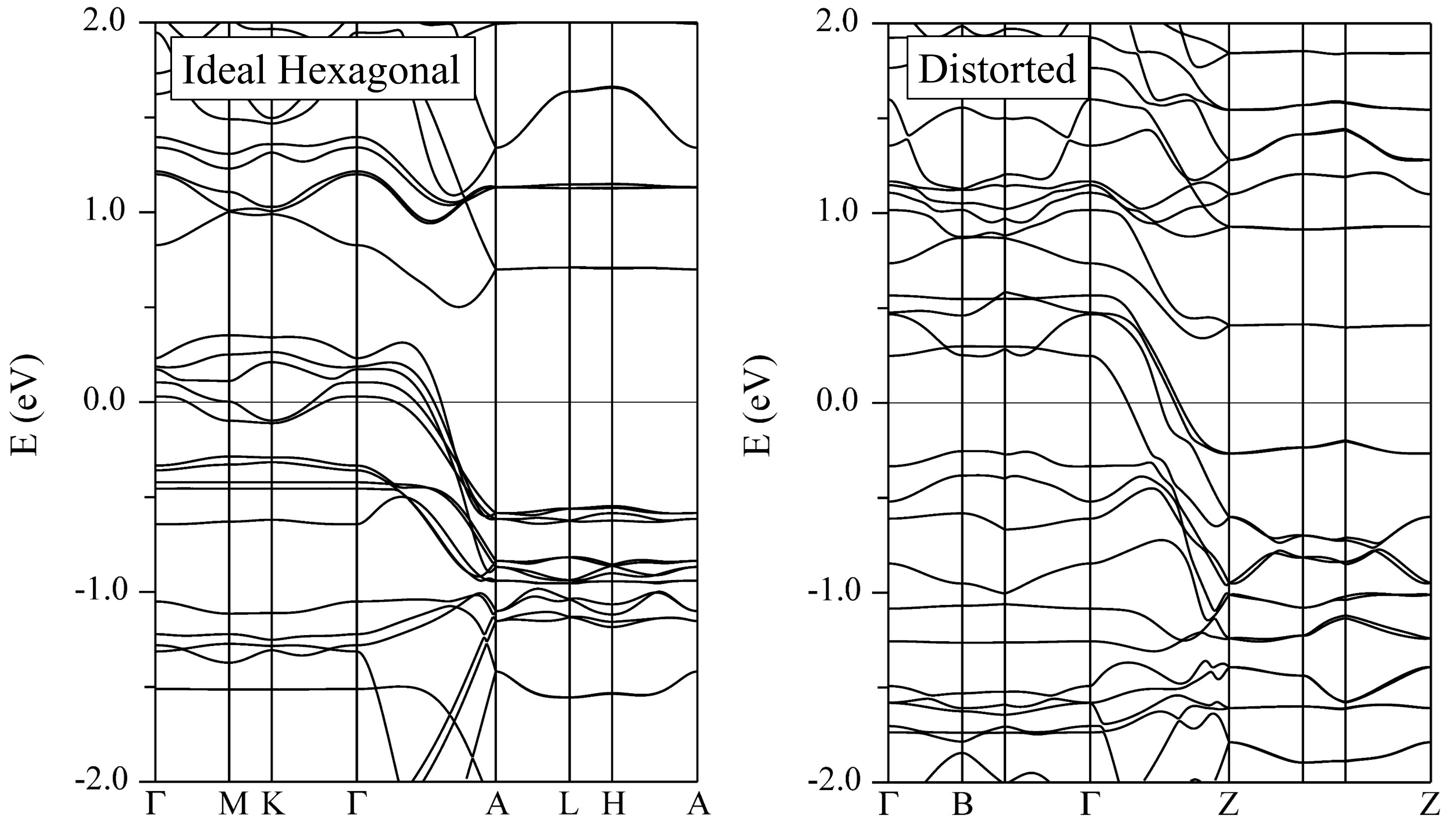}
\caption{Band structures of the ideal hexagonal and distorted structures
of KCr$_3$As$_3$, showing equivalent paths through the zone. Non-symmetry
points for the distorted structure are not labeled.}
\label{bands}
\end{figure}

We did magnetic calculations considering the collinear magnetic structures
described by Cao and co-workers. \cite{KCr3As3-cal} including the
doubling of the unit cell along the $c$-axis, for both the
distorted and undistorted structures. For the undistorted structure, our
results are qualitatively similar to theirs (note that Cao and
co-workers used slightly different hexagonal space group for many of the
calculations that they report).
We find a magnetic ground state with
both the PBE GGA and the local spin density approximation. The lowest
energy state in our calculation using the experimental space group ($P6_3/m$),
\cite{KCr3As3,KCr3As3-5K}
consists of Cr layers with alternating spin
along the $c$-axis.
This is a $\Gamma$-point instability, since the
unit cell contains two layers.

Magnetic materials are often classified according to the extent to
which they are itinerant or local moment in nature.
\cite{moriya,svanidze}
In the local moment limit, stable moments are present on the ions due to
intra-atomic interactions. These are subject to inter-site interactions
that provide the magnetism through ordering of the moments. These
can often be described by the Heisenberg or related spin models.
Related to KCr$_3$As$_3$.
BaMn$_2$As$_2$ is a material close to this limit.
\cite{johnston}
The opposite, itinerant limit, is exemplified by materials such as
ZrZn$_2$, Y$_2$Ni$_7$ and to a lesser extent elemental Ni.
\cite{moriya}
In this limit, there are not stable moments, but instead ordering
arises from conduction electrons, and in fact magnetic states do not
exist for some ordering patterns.
This is seen for example from merging
of spin-wave excitations seen by neutron scattering into a Stoner 
continuum at high wavevectors. \cite{mook}
Both of these types of magnetism can be described by DFT calculations,
as examplified by work on BaMn$_2$As$_2$, Y$_2$Ni$_7$ and other
materials.
\cite{an,honer,y2ni7}

We find that the magnetic
instabilities are itinerant in nature since the magnetic moments depend
strongly on the particular order and the energy differences between
different magnetic orders are on the same scale as that of ordered states
relative to the non-spin-polarized energy.
For example, the ferromagnetic solution
has moments that are less than half of the lowest energy antiferromagnetic
solution, i.e. 0.21 $\mu_B$/Cr vs. 0.53 $\mu_B$/Cr, as measured by
the spin moment in an LAPW sphere, radius 2.25 bohr.
The energy,
with respect to the non-spin-polarized solution,
of the ferromagnetic state is -8 meV,
while the antiferromagnetic is -19 meV both on a per formula unit basis.
Thus in its observed hexagonal structure, KCr$_3$As$_3$ is
predicted to be an itinerant antiferromagnet. The non-magnetic ground
state observed would then require strong spin fluctuations, as may
occur near a quantum critical point in analogy with unconventional
superconductors such as the Fe-based materials.
\cite{mazin-mag,bondino}
However, with the structure distortion the situation is completely
different. In particular, we find no magnetic instabilities in the
distorted structure.

The Fermi surfaces for the hexagonal and distorted structures are
also very different as shown
in Fig. \ref{fermi}. 
In particular, the
prominent three-dimensional Fermi surface in the undistorted structure
is completely removed by the distortion. Left are four very one-dimensional
Fermi surface sheets. All of these contain holes at the zone center (electrons
at $k_z$=0.5). The electron-like volumes enclosed are 0.709, 0.510,
0.426 and 0.355 in terms of the Brillouin zone volume. The corresponding
nesting vectors of these 1D sheets
are along $k_z$ and are 0.291, 0.490, 0.426 and 0.355
in units of $2\pi/c$.
The band structures are shown in Fig. \ref{bands}. As seen, the change in Fermi
surface is related to an upward shift of the bands giving rise to the 3D
sections.

\section{Discussion and Conclusions}

To summarize, we find strong structural instabilities of the reported
hexagonal structure of KCr$_3$As$_3$. These lead to a much more one-dimensional
Fermi surface structure, a depletion of $N(E_F)$ and a suppression of
magnetism.

The structure calculated here is an ordered structure with distorted
Cr$_3$As$_3$ tubes. Considering the experimental fact that the crystal
structure of KCr$_3$As$_3$ is reported to be hexagonal, and
the phonon dispersions that suggest low coherence of the distortion
between tubes,
the likely state at low temperature consists of distorted tubes with
inter-tube disorder. This is similar to what is found in K$_2$Cr$_3$As$_3$.
The present results show a strong coupling between the distortions
and the three-dimensional Fermi surface, while the one-dimensional
sheets are left. It would therefore seem likely that disordered 
distortions would also destroy the three-dimensional Fermi surface.
It would be important to verify this experimentally, by direct
measurements, e.g. with ARPES, and also by transport measurements, 
where for example the conductivity anisotropy could be used to 
probe whether or not the three-dimensional Fermi surface remains.
It will also be of importance to perform detailed experimental
structural studies,
such as those reported for K$_2$Cr$_3$As$_3$ to check whether the
actual ground state structure is distorted and to determine the
details.

Finally, we note that the results have implications for the superconductivity.
While there are no magnetic instabilities found for the distorted structure,
it might be supposed that the nesting of one-dimensional
Fermi surface sheets would provide spin fluctuations
near the nesting vectors. However, these would not provide a mechanism
for superconductivity. The reason is that according to the spin fluctuation
theory of Berk and Schrieffer, \cite{berk}
spin fluctuations are repulsive for a singlet state and attractive for
a triplet. For 1D Fermi surfaces at $k_z$ and $-k_z$ a repulsive (singlet)
interaction would favor opposite sign order parameter on the two sheets, 
which is not a singlet state, while an attractive interaction (triplet)
would conversely favor same sign (singlet) order parameter, which is not
a triplet.

Thus the Pauli exclusion principle prevents effective pairing on
such 1D sheets based on nesting related spin fluctuations.
On the other hand, electron phonon interactions are attractive
for both singlet and triplet pairing.
Nesting of quasi-one-dimensional sheets is expected
to lead to strong electron phonon coupling, which could support a conventional
singlet $s$ wave state. In this regard, it will be of interest to
search for Kohn anomalies in the phonon dispersions using neutron
scattering measurements.

\begin{acknowledgments}

We thank Keith Taddei for helpful discussions.
This work was supported by the U.S. Department of Energy, Office of Science,
Office of Basic Energy Sciences, Award Number DE-SC0019114.
Support for work at Shanghai University was provided by the
National Natural Science Foundation of China
(Grants 51672171 and 51861145315),
the National Key Basic Research Program of China (Grant 2015CB921600),
the fund of the State Key Laboratory of Solidification Processing in NWPU
(SKLSP201703) and the Fok Ying Tung Education Foundation. 
GX is grateful
for support from the China Scholarship Council.

\end{acknowledgments}

\bibliography{reference}

\end{document}